\documentclass[twocolumn,preprintnumbers,amsmath,superscriptaddress,amssymb,aps]{revtex4-1}

\usepackage{graphicx}
\usepackage{dcolumn}
\usepackage{bm}
\usepackage{amsmath}
\usepackage{amssymb}
\usepackage{graphicx}
\usepackage{indentfirst}
\usepackage{booktabs}
\usepackage{multirow}
\usepackage{colortbl}

\selectfont

\begin{document}
\title{Ferroelectric tuning of the valley polarized metal-semiconductor transition in Mn$_2$P$_2$S$_3$Se$_3$/Sc$_2$CO$_2$ van der Waals heterostructures and application to nonlinear Hall effect devices}

\author{Hanbo Sun}
\thanks{These authors contributed equally to this work.}
\address{State Key Laboratory for Mechanical Behavior of Materials, School of Materials Science and Engineering, Xi'an Jiaotong University, Xi'an, Shaanxi, 710049, People's Republic of China}
\author{Yewei Ren}
\thanks{These authors contributed equally to this work.}
\address{State Key Laboratory for Mechanical Behavior of Materials, School of Materials Science and Engineering, Xi'an Jiaotong University, Xi'an, Shaanxi, 710049, People's Republic of China}
\author{Chao Wu}
\address{State Key Laboratory for Mechanical Behavior of Materials, School of Materials Science and Engineering, Xi'an Jiaotong University, Xi'an, Shaanxi, 710049, People's Republic of China}
\author{Pengqiang Dong}
\address{State Key Laboratory for Mechanical Behavior of Materials, School of Materials Science and Engineering, Xi'an Jiaotong University, Xi'an, Shaanxi, 710049, People's Republic of China}
\author{Weixi Zhang}
\email{zhangwwxx@sina.com}
\address{Department of Physics and Electronic Engineering, Tongren University, Tongren 554300, People's Republic of China}
\author{Yin-Zhong Wu}
\address{School of Physical Science and Technology, Suzhou University of Science and Technology, Suzhou 215009, People's Republic of China}
\author{Ping Li}
\email{pli@xjtu.edu.cn}
\address{State Key Laboratory for Mechanical Behavior of Materials, School of Materials Science and Engineering, Xi'an Jiaotong University, Xi'an, Shaanxi, 710049, People's Republic of China}
\address{State Key Laboratory for Surface Physics and Department of Physics, Fudan University, Shanghai, 200433, People's Republic of China}
\address{State Key Laboratory of Silicon and Advanced Semiconductor Materials, Zhejiang University, Hangzhou, 310027, People's Republic of China}

\date{\today}

\begin{abstract}
In order to promote the development of the next generation of nano-spintronic devices, it is of great significance to tune the freedom of valley in two-dimensional (2D) materials. Here, we propose a mechanism for manipulating the valley and nonlinear Hall effect by the 2D ferroelectric substrate. The monolayer Mn$_2$P$_2$S$_3$Se$_3$ is a robust antiferromagnetic valley polarized semiconductor. Importantly, the valley polarized metal-semiconductor phase transition of Mn$_2$P$_2$S$_3$Se$_3$ can be effectively tuned by switching the ferroelectric polarization of Sc$_2$CO$_2$. We reveal the microscopic mechanism of phase transition, which origins from the charge transfer and band alignment. Additionally, we find that transformed polarization direction of Sc$_2$CO$_2$ flexibly manipulate the Berry curvature dipole. Based on this discovery, we present the detection valley polarized metal-semiconductor transition by the nonlinear Hall effect devices. These findings not only offer a scheme to tune the valley degree of freedom, but also provide promising platform to design the nonlinear Hall effect devices.
\end{abstract}

\maketitle
\section{Introduction}
Limited by the storage wall problem and von Neumann bottleneck, traditional silicon-based memory cannot meet the demands of massive data processing and storage \cite{1,2}. Therefore, revolutionary memory technologies with the low power consumption, ultra-fast speed, ultra-high capacity, and non-volatile based on alternative mechanisms, structures, and materials are highly profiled \cite{3,4,5,6}. Among the potential candidates, the booming development of two-dimensional (2D) materials and their heterojunctions provides promising opportunities due to their ideal atomic-level flatness without dangling bonds, unique electronic properties, and highly adjustability \cite{7,8,9,10}.

Recently, the concept of ferrovalley materials with spontaneous valley polarization was proposed by Duan $\emph{et al.}$ \cite{11}. Ferrovalley is a combination of valley degree of freedom and ferro-order \cite{12,13,14,15,16}. Consequently, it possesses not only valley degree of freedom features, but also magnetic or ferroelectric characteristics. It has potential applications in information storage, processing, and transmission based on coding and probing of valley (spin) degree of freedom. And it can be used to construct valley (spin) valves, and valley (spin) filters, valleytronic (spintronic) devices \cite{17,18,19,20}. Until now, a large amount of ferrovalley materials have been reported, such as FeX$_2$ (X = Cl, Br) \cite{21,22}, Cr$_2$Se$_3$ \cite{23}, XY (X = K, Rb, Cs; Y = N, P, As, Sb, Bi) \cite{24}, YI$_2$ \cite{25}, VSiXN$_4$ (X = C, Si, Ge, Sn, Pb) \cite{26}, Fe$_2$CF$_2$ \cite{27}, and so on. Nevertheless, a large number of ferrovalley materials are predicted, but how to realize the non-volatile control valley that is the key to achieve the valleytronic devices.

Moreover, the nonlinear Hall effect (NHE) has been recently discovered \cite{28,29}. Unlike the linear anomalous Hall effect that the transverse Hall currents is linearly proportional to the longitudinal driving electric field, where only presents in magnetic systems. However, the transverse Hall voltage of the NHE is quadratic with the driving current, requiring the breaking of spatial inversion ($\emph{P}$) symmetry \cite{30,31,32,33}. If the system further breaks the time-reversal symmetry ($\emph{T}$), how does the NHE change? More importantly, whether 2D ferroelectric substrate can effectively tune the nonlinear Hall conductance?

In this work, we propose that the valley polarization and nonlinear Hall conductance of Janus Mn$_2$P$_2$S$_3$Se$_3$ can be tuned by 2D ferroelectric substrate Sc$_2$CO$_2$. Based on the density functional theory (DFT), we comprehensively investigate the magnetic ground state, electronic properties, valleytronic properties, and Berry curvature dipole (BCD) for the Mn$_2$P$_2$S$_3$Se$_3$/Sc$_2$CO$_2$ van der Waals (vdW) heterostructure. We find that the valley polarized metal-semiconductor transition of Mn$_2$P$_2$S$_3$Se$_3$ can be effectively realized by altering the ferroelectric polarization of Sc$_2$CO$_2$. Furthermore, switching the electric polarization direction can induce the large BCDs. Our findings provide valuable guidances for the future development of valleytronic nano switches and the NHE devices.

\section{COMPUTATIONAL METHODS}
All calculations are implemented based on the framework of the DFT using the Vienna $Ab$ $initio$ simulation package (VASP) \cite{34,35}. The generalized gradient approximation (GGA) with the Perdew-Burke-Ernzerhof (PBE) functional is used to describe the exchange-correlation energy \cite{36}. The plane-wave basis with a kinetic energy cutoff of 500 eV is employed. A vacuum of 25 $\rm \AA$ is set along the $\emph{c}$-axis, to avoid the interaction between the sheet and its periodic images. The convergence criteria of the total energy and the force are set to the 10$^{-6}$ eV and -0.01 eV/$\rm \AA$, respectively. The zero damping DFT-D3 method of Grimme is considered for the vdW correction in Mn$_2$P$_2$S$_3$Se$_3$/Sc$_2$CO$_2$ heterostructure \cite{37}. To describe strongly correlated 3$\emph{d}$ electrons of Mn, the GGA+U method is applied with the effective U value (U$_{eff}$ = U - J) of 4 eV. To study the dynamical stability, the phonon spectra are calculated by the PHONOPY code using a 4$\times$4$\times$1 supercell \cite{38}. The maximally localized Wannier functions (MLWFs) are performed to construct an effective tight-binding Hamiltonian to investigate the Berry curvature and BCD by Wannier90 code \cite{39,40}. In the BCD calculations, a $\textbf{k}$ mesh grids of 1000$\times$1000$\times$1 is adopted to ensure a convergent of the results.

\section{RESULTS AND DISCUSSION }	
\subsection{The material model of Mn$_2$P$_2$S$_3$Se$_3$/Sc$_2$CO$_2$ van der Waals heterostructures}
Before building the atomic models of the Mn$_2$P$_2$S$_3$Se$_3$/Sc$_2$CO$_2$ vdW multiferroic heterostructures, we first investigate the crystal structures of monolayer Mn$_2$P$_2$S$_3$Se$_3$ and Sc$_2$CO$_2$, as shown in Fig. 1(a, b) and Fig. 1(d, e). The monolayer Janus Mn$_2$P$_2$S$_3$Se$_3$ shows a hexagonal lattice with the point group of C$_{3v}$, which it naturally breaks the symmetry of $\emph{P}$ \cite{41}. Moreover, the asymmetric displacement of the C sublayer in relative to the Sc sublayer also results in the breaking the $\emph{P}$ for monolayer Sc$_2$CO$_2$, inducing the out-of-plane ferroelectric polarization \cite{42}. After the structural relaxation, the in-plane lattice constants of monolayer Mn$_2$P$_2$S$_3$Se$_3$ and Sc$_2$CO$_2$ are 6.25 $\rm \AA$ and 3.42 $\rm \AA$, respectively.

\begin{figure}[htb]
\begin{center}
\includegraphics[angle=0,width=1.0\linewidth]{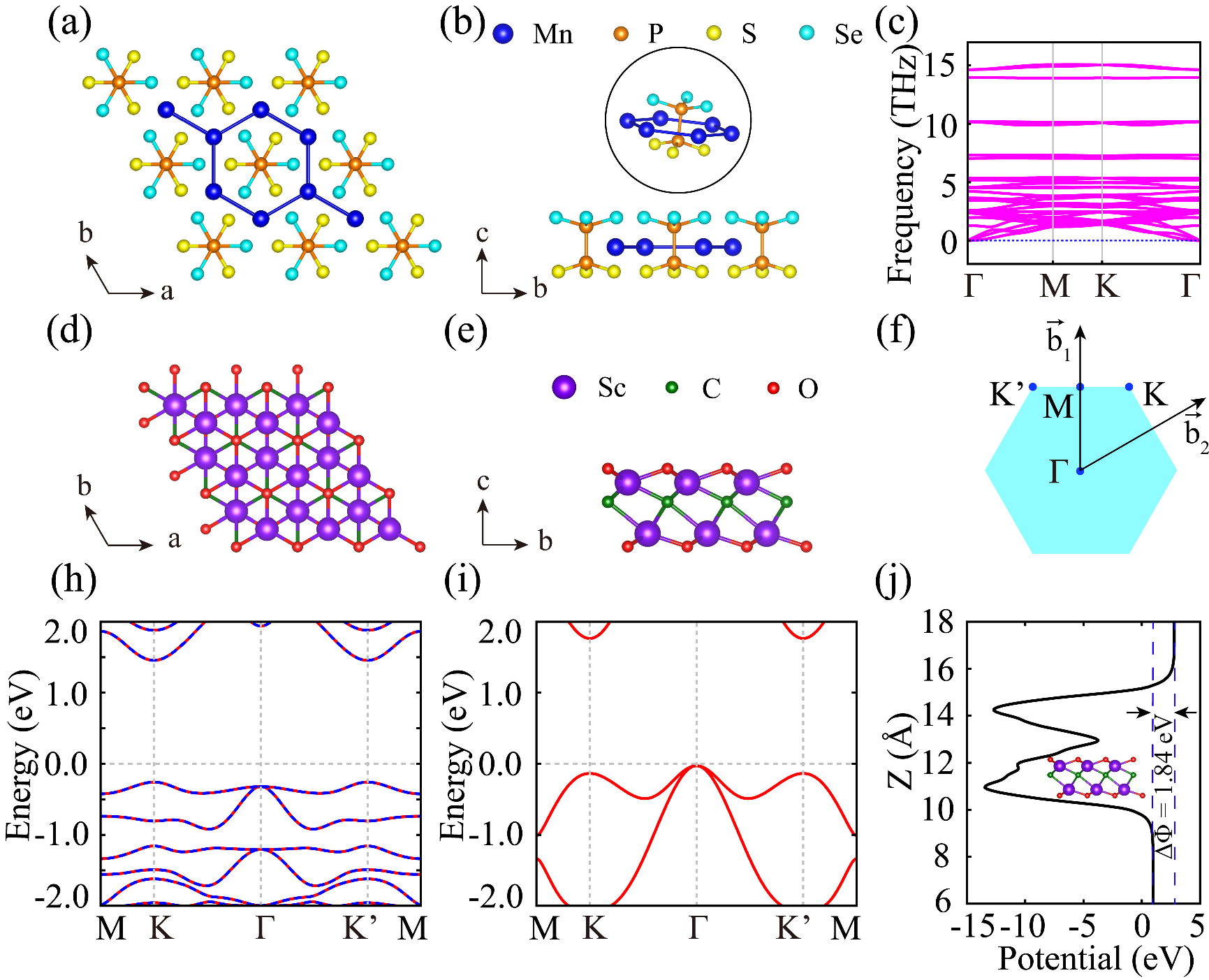}
\caption{(a, b) The top and side views of the crystal structure for monolayer Mn$_2$P$_2$S$_3$Se$_3$, respectively. The dark blue, orange, yellow, and light blue balls represent Mn, P, S, and Se elements, respectively. (c) The calculated phonon dispersion curves of monolayer Mn$_2$P$_2$S$_3$Se$_3$ along the high-symmetry lines. (d, e) The top and side views of the crystal structure for monolayer Sc$_2$CO$_2$, respectively. The purple, green, and red balls represent Sc, C, and O elements, respectively. (f) The Brillouin zone (BZ) of the honeycomb lattice with the reciprocal lattice vectors $\vec{b}_1$ and $\vec{b}_2$. The $\Gamma$, K, K', and M are high-symmetry points in the BZ. (h) Spin-polarized band structure of monolayer Mn$_2$P$_2$S$_3$Se$_3$. The solid red line and dashed blue line represent spin up and spin down bands, respectively. (i) Band structure of monolayer Sc$_2$CO$_2$ without considering the SOC. (j) The plane averaged electrostatic potential of monolayer Sc$_2$CO$_2$, in which $\Delta \Phi$ represents the potential difference.
}
\end{center}
\end{figure}

Moreover, we also study the properties of monolayer Mn$_2$P$_2$S$_3$Se$_3$. Firstly, we evaluate the dynamic stability by the phonon dispersion spectrum. As shown in Fig. 1(c), the absence of imaginary modes along the high-symmetry lines confirms the dynamical stability of monolayer Mn$_2$P$_2$S$_3$Se$_3$. Then, to determine that the Coulomb repulsion U doesn't effect on the magnetic ground state, we calculate the ferromagnetic (FM) and antiferromagnetic (AFM) states between 1 eV and 5 eV. As listed in Table SI, the total energy of the AFM state is always lower than that of FM state \cite{43}. It indicates that the AFM state is robust. The magnetic anisotropy energy (MAE) is the basis of investigating the properties of magnetic materials. The MAE is defined as the total energy difference between the magnetic moment along in-plane (E$_{100}$) and out-of-plane (E$_{001}$). As listed in Table SI, the Hubbard U has little effect on the MAE \cite{43}. Both are negative, meaning that the direction of easy magnetization is in-plane. With such a small MAE, it is easily tuned out-of-plane by a small external magnetic field \cite{12}. In previous reports, the Coulomb repulsion U is often taken to be 4 eV for the Mn atom \cite{44,45}. Hence, we choose U$_{\rm eff}$ = 4 eV to investigate all the properties that follow.

\begin{table*}[htb]
\caption{
The binding energies ($\emph{E}$$_b$) and interlayer distance ($\emph{d}$) of different stacking configurations for Mn$_2$P$_2$S$_3$Se$_3$/Sc$_2$CO$_2$ heterostructures.
}
\begin{tabular*}{0.62\linewidth}{cccccccc}
	\hline
            & \emph{}                   & Mn$_2$P$_2$S$_3$Se$_3$/Sc$_2$CO$_2$($\uparrow$) \quad   & Mn$_2$P$_2$S$_3$Se$_3$/Sc$_2$CO$_2$($\downarrow$)               \\
	        & Configuration             & \emph{E$_b$} (meV)  \quad   \emph{d} (${\rm \AA}$)      & \emph{E$_b$} (meV)  \quad   \emph{d} (${\rm \AA}$)              \\
	\hline
	        & top-O                     & -286.86             \quad   2.58                        & -254.24             \quad   2.71                                \\
S-surface 	& top-Sc                    & -226.90             \quad   2.88                        & -213.21             \quad   2.96                                \\
            & hollow                    & -180.01             \quad   3.26                        & -161.72             \quad   3.38                                \\
	\hline
            & top-O                     & -277.91             \quad   2.73                        & -256.86             \quad   2.80                                \\
Se-surface  & top-Sc                    & -239.64             \quad   2.96                        & -226.97             \quad   3.05                                \\
            & hollow                    & -191.99             \quad   3.34                        & -176.09             \quad   3.45                                \\
	\hline
\end{tabular*}
\end{table*}	

For the electronic properties, Fig. 1(h) and 1(i) exhibit the band structures of monolayer Mn$_2$P$_2$S$_3$Se$_3$ and Sc$_2$CO$_2$ without considering the spin-orbit coupling (SOC), respectively. Evidently, monolayer Mn$_2$P$_2$S$_3$Se$_3$ appears a direct band gap semiconductor with the band gap of 1.71 eV located at the K/K' point. Specifically, spin up and spin down channels are degenerate due to the AFM magnetic ground state. When the SOC is included, as shown in Fig. S1 \cite{43}, the degeneracy between K and K' valleys disappears. Simultaneously, the conduction band minimum (CBM) is induced the valley splitting of 25.80 meV. Monolayer Sc$_2$CO$_2$ shows a insulator with an indirect band gap of 1.79 eV. In addition, the out-of-plane polarized of Sc$_2$CO$_2$ introduces an electrostatic potential difference 1.84 eV between the two surfaces of Sc$_2$CO$_2$, as shown in Fig. 1(j).

\begin{figure}[htb]
\begin{center}
\includegraphics[angle=0,width=1.0\linewidth]{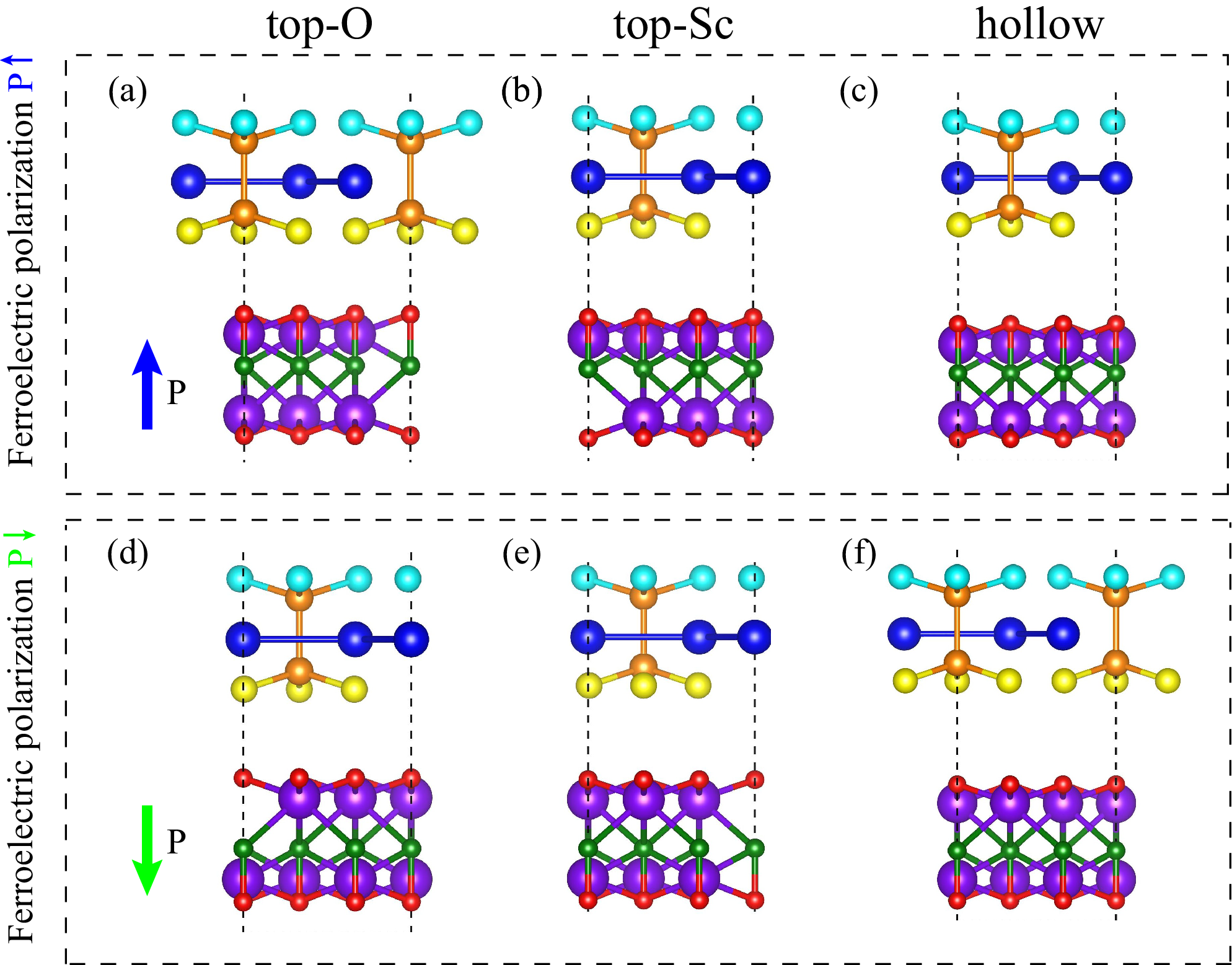}
\caption{(a-f) Side views of the Mn$_2$P$_2$S$_3$Se$_3$/Sc$_2$CO$_2$ heterostructure with diverse stacking configurations under different Sc$_2$CO$_2$ polarization states. The interface contact is the contact between S-surface of Mn$_2$P$_2$S$_3$Se$_3$ and Sc$_2$CO$_2$. For the top-O (a, d), top-Sc (b, e), and hollow (c, f) configurations, (a-c) and (d-f) are shown the P$\uparrow$ and the P$\downarrow$ states, respectively.
}
\end{center}
\end{figure}

Note that the lattice mismatches is 5.6 $\%$ for the 1$\times$1 Mn$_2$P$_2$S$_3$Se$_3$ matching to the $\sqrt{3}$$\times$$\sqrt{3}$ Sc$_2$CO$_2$. Considering that the lattice mismatch rate is too large, we optimized the lattice constant of the heterostructure to 5.98 $\rm \AA$. We have considered six typical alignments, i.e., the bottom layer S/Se atom of Mn$_2$P$_2$S$_3$Se$_3$ being in the top and hollow positions of the Sc$_2$CO$_2$ honeycomb lattice, respectively. As shown in Fig. 2 and Fig. S2 \cite{43}, we named it top-O, top-Sc, and hollow configurations. After extensive geometry optimizations, we find that the layer spacing is 2.58 $\sim$ 3.45 ${\rm \AA}$, as listed in Table I. It indicates the nature of vdW interaction between Mn$_2$P$_2$S$_3$Se$_3$ and Sc$_2$CO$_2$. Besides, to evaluate the stability of these configurations, we calculate the binding energy E$_b$. E$_b$ defined as E$_b$ = (E$_{\rm total}$ - E$_{\rm Mn_2P_2S_3Se_3}$ - E$_{\rm Sc_2CO_2}$)/N$_{\rm S/Se}$, where E$_{\rm total}$, E$_{\rm Mn_2P_2S_3Se_3}$, and E$_{\rm Sc_2CO_2}$ are the total energies of Mn$_2$P$_2$S$_3$Se$_3$/Sc$_2$CO$_2$, Mn$_2$P$_2$S$_3$Se$_3$, and Sc$_2$CO$_2$, respectively. N$_{\rm S/Se}$ is the number of S/Se atoms at the interface. The E$_b$ is -161.72 $\sim$ -286.86 meV, meaning that the configurations are stable and likely to be synthesized experimentally. Importantly, whether the S-surface or Se-surface polarization is upward or downward, the most stable is the top-O configuration.

\subsection{Ferroelectric tuning valley polarized metal-semiconductor transition and mechanism}
To confirm the magnetic ground state of Mn$_2$P$_2$S$_3$Se$_3$/Sc$_2$CO$_2$ heterostructure, two possible magnetic configurations including the FM and AFM are considered. The relative energies (E$_{\rm FM}$ - E$_{\rm AFM}$) of the FM and AFM phases, are listed in Table SII \cite{43}. The energy difference is always positive for all configurations, indicating the magnetic ground state is AFM state in Mn$_2$P$_2$S$_3$Se$_3$/Sc$_2$CO$_2$ heterostructure. Furthermore, the MAE of heterostructure decreases compared with that of monolayer, as shown in Table SIII \cite{43}.

\begin{figure}[htb]
\begin{center}
\includegraphics[angle=0,width=1.0\linewidth]{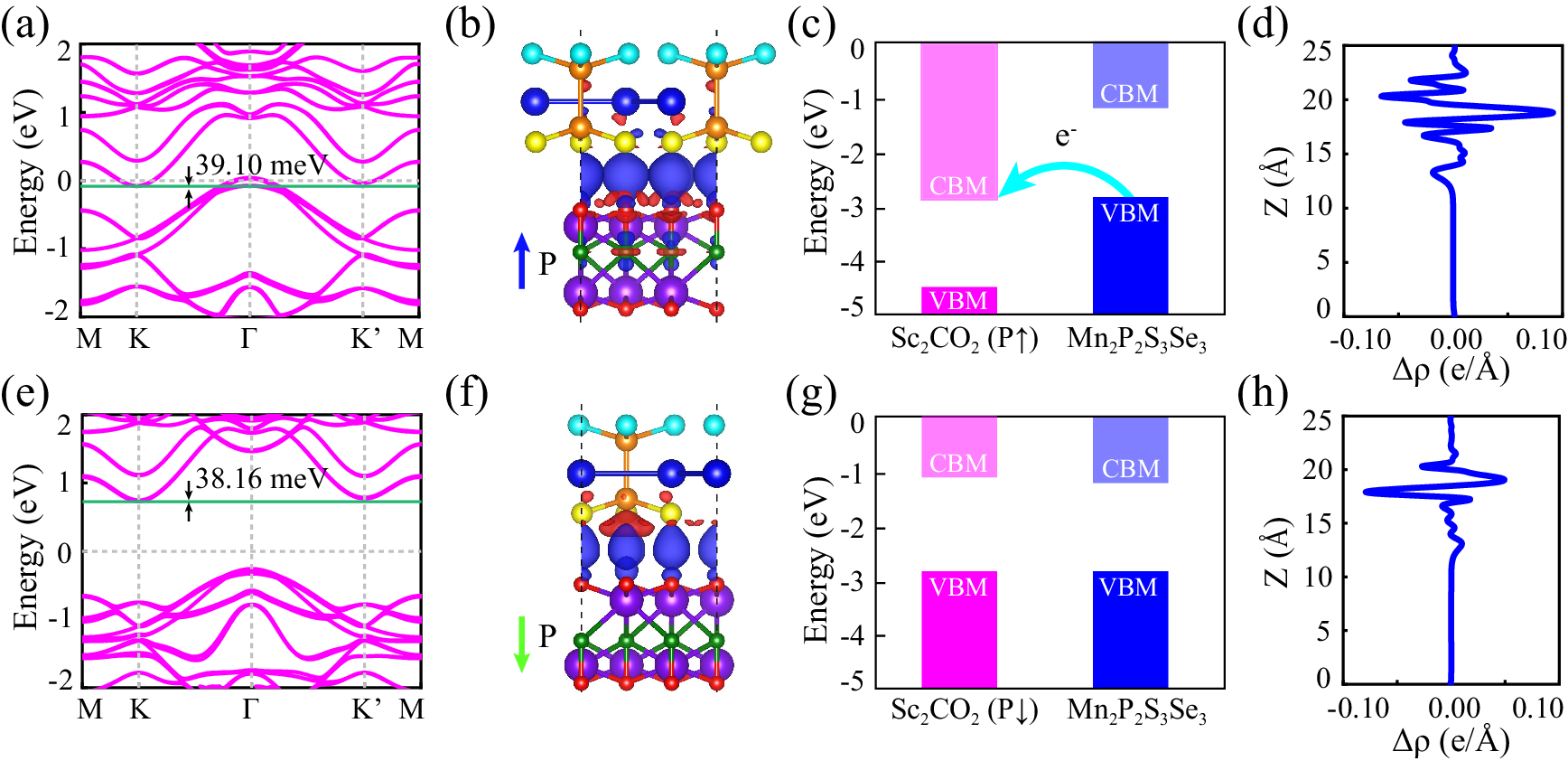}
\caption{The top-O configuration Mn$_2$P$_2$S$_3$Se$_3$/Sc$_2$CO$_2$ heterostructures for the S-surface. (a, e) Band structures with considering the SOC under the P$\uparrow$ (a) and the P$\downarrow$ (e) states. (b, f) The 3D charge density difference under the P$\uparrow$ (b) and the P$\downarrow$ (f) states. (c, g) Band alignments of the P$\uparrow$ (c) and the P$\downarrow$ (g) states. (d, h) Plane averaged charge density difference for the P$\uparrow$ (d) and the P$\downarrow$ (h) states. The conduction band valley splitting is indicated by the green shading.
}
\end{center}
\end{figure}

Primarily, we investigate the S-surface. As shown in Fig. S3, when the SOC is switched off, spin up and spin down channels are still degenerate \cite{43}. In the Mn$_2$P$_2$S$_3$Se$_3$/Sc$_2$CO$_2$ P$\uparrow$ case, they exhibit metallic properties except hollow configuration (see Fig. S3(a-c)) \cite{43}. More interestingly, the greater the binding energy between Mn$_2$P$_2$S$_3$Se$_3$ and Sc$_2$CO$_2$, the stronger the metallic property of the system, as listed in Table I. However, when the ferroelectric polarization of Sc$_2$CO$_2$ switches to the downward, the band gap is opened becoming a semiconductor (see Fig. S3(d-f)) \cite{43}. When the SOC is included, as shown in Fig. 3(a, e) and Fig. S4, we find that the CBM of K and K' occur valley splitting with the 35.41 $\sim$ 39.10 meV. At this moment, the Mn$_2$P$_2$S$_3$Se$_3$/Sc$_2$CO$_2$ P$\uparrow$ become valley polarized metal phase. However, the band structures of Mn$_2$P$_2$S$_3$Se$_3$/Sc$_2$CO$_2$ P$\downarrow$ exhibit valley polarized semiconductor characteristics.

In order to understand the origin of valley splitting for Mn$_2$P$_2$S$_3$Se$_3$/Sc$_2$CO$_2$ heterostructure, we calculated the orbital-resolved band structure for the Mn atoms $\emph{d}$ orbital, as shown in Fig. S5 \cite{43}. The CBM bands of K and K' points are mainly contributed by Mn d$_{xy}$/d$_{x^2-y^2}$ orbitals for all configurations, while the valence band maximum (VBM) bands of K and K' points for the P$\uparrow$ and P$\downarrow$ states are dominated by d$_{z^2}$ and d$_{xz}$/d$_{yz}$ orbitals, respectively. Then, we built an effective Hamiltonian model based on the SOC effect as a perturbation term. The SOC Hamiltonian can be described as
\begin{equation}
\hat{H}_{SOC} = \lambda \hat{S} \cdot \hat{L} = \hat{H}_{SOC}^{0} + \hat{H}_{SOC}^{1},
\end{equation}
where $\hat{L}$ and $\hat{S}$ are orbital angular and spin angular operators, respectively. $\hat{H}_{SOC}^{0}$ and $\hat{H}_{SOC}^{1}$ denote the interaction between the same spin states and between opposite spin states, respectively. Since the valley splitting mainly occurs at the CBM bands, for Mn$_2$P$_2$S$_3$Se$_3$/Sc$_2$CO$_2$ heterostructure, we focus on the mechanism of the CBM valley splitting. Although the CBM is composed of spin up and spin down bands, they are degenerate. Therefore, we only consider the term $\hat{H}_{SOC}^{0}$ and ignore the term $\hat{H}_{SOC}^{1}$. Moreover, $\hat{H}_{SOC}^{0}$ can be written in polar angles
\begin{equation}
\hat{H}_{SOC}^{0} = \lambda \hat{S}_{z'}(\hat{L}_zcos\theta + \frac{1}{2}\hat{L}_+e^{-i\phi}sin\theta + \frac{1}{2}\hat{L}_-e^{+i\phi}sin\theta),
\end{equation}
In the out-of-plane magnetization case, $\theta$ = $\phi$ = 0$^ \circ$, then the $\hat{H}_{SOC}^{0}$ term can be simplified as
\begin{equation}
\hat{H}_{SOC}^{0} = \lambda \hat{S}_{z} \hat{L}_z,
\end{equation}

Taking into account the orbital contribution around the CBM valleys and the C$_{3v}$ symmetry, we taken $|$$\psi$$_c$$^{\tau}$$\rangle$=$\frac{1}{\sqrt{2}}$($|$d$_{xy}$$\rangle$+i$\tau$$|$d$_{x2-y2}$$\rangle$)$\otimes$$|$$\uparrow$$\rangle$ as the orbital basis for the CBM, where $\tau$ = $\pm$1 represent the valley index corresponding to $\rm K/\rm K'$. The energy levels of the valleys for the CBM can be written as E$_c$$^ \tau$ = $\langle$ $\psi$$_c$$^ \tau$ $|$ $\hat{H}$$_{SOC}^{0}$ $|$ $\psi$$_c$$^ \tau$ $\rangle$. Then, the valley polarization can be expressed as
\begin{equation}
E_{c}^{K} - E_{c}^{K'} = i \langle d_{xy} | \hat{H}_{SOC}^{0} | d_{x2-y2} \rangle - i \langle d_{x2-y2} | \hat{H}_{SOC}^{0} | d_{xy} \rangle \approx 4\beta,
\end{equation}
where the $\hat{L}_z|d_{xy} \rangle$ = -2i$\hbar$$|d_{x2-y2} \rangle$, $\hat{L}_z|d_{x2-y2} \rangle$ = 2i$\hbar$$|d_{xy} \rangle$, and $\beta = \lambda \langle d_{x2-y2} |\hat{S}_{z'}| d_{x2-y2} \rangle$. The analytical result proves that the valley degeneracy split is consistent with our DFT calculations ($E_{c}^{K}$ - $E_{c}^{K'}$ = 35.41 $\sim$ 39.10 meV).

To understand the physics mechanism of the Sc$_2$CO$_2$ polarization switching driven valley polarized metal-semiconductor transition, we calculated the three-dimensional (3D) charge density difference and plane averaged charge density difference ($\Delta \rho$), as shown in Fig. 3(b, f, d, h). The negative and positive values (blue and red areas) denote electron depletion and accumulation, respectively. Obviously, the Mn$_2$P$_2$S$_3$Se$_3$/Sc$_2$CO$_2$ P$\uparrow$ system shows a notable charge transfer from Mn$_2$P$_2$S$_3$Se$_3$ to Sc$_2$CO$_2$. Conversely, in the Mn$_2$P$_2$S$_3$Se$_3$/Sc$_2$CO$_2$ P$\downarrow$ case, the Mn$_2$P$_2$S$_3$Se$_3$ serves as an acceptor and accumulates electrons, whereas the Sc$_2$CO$_2$ acts as a donor and deplete electrons. In addition, we calculated the band alignments of Mn$_2$P$_2$S$_3$Se$_3$/Sc$_2$CO$_2$ heterostructure. As shown in Fig. 3(c, g), the VBM and CBM are acquired from freestanding Mn$_2$P$_2$S$_3$Se$_3$ and Sc$_2$CO$_2$. In the Sc$_2$CO$_2$ P$\uparrow$ case, the CBM of Sc$_2$CO$_2$ is lower than the VBM of Mn$_2$P$_2$S$_3$Se$_3$, electrons tend to migrate from Mn$_2$P$_2$S$_3$Se$_3$ to Sc$_2$CO$_2$. Due to the inherent electric field, there is a potential difference of 1.84 eV between the two sides of Sc$_2$CO$_2$, resulting in obviously different band shifts in Mn$_2$P$_2$S$_3$Se$_3$ when touching with opposite sides. When the Sc$_2$CO$_2$ P is $\downarrow$ case, the VBM of Sc$_2$CO$_2$ slightly above the VBM of Mn$_2$P$_2$S$_3$Se$_3$, leading to electron transfers from Sc$_2$CO$_2$ to Mn$_2$P$_2$S$_3$Se$_3$. Importantly, it is exactly consistent with the charge transfer.

For the Se-surface case, its result is completely similar to the S-surface. In the absence of SOC, spin up and spin down bands are degenerate, as shown in Fig. S6 \cite{43}. The system shows the metallic properties besides hollow configuration for the Mn$_2$P$_2$S$_3$Se$_3$/Sc$_2$CO$_2$ P$\uparrow$ (see Fig. S6(a-c)) \cite{43}, while the Mn$_2$P$_2$S$_3$Se$_3$/Sc$_2$CO$_2$ P$\downarrow$ retains the semiconductor (see Fig. S6(d-f)) \cite{43}. When the SOC is switched on, as shown in Fig. 4(a, e) and Fig. S7 \cite{43}, the CBM of K and K' appear the 34.16 $\sim$ 39.53 meV valley splitting. Moreover, Fig. S8 exhibits the calculated projected band structures for the Mn atomic $\emph{d}$ orbital \cite{43}. The CBM dominated from d$_{xy}$/d$_{x^2-y^2}$ orbital of Mn atom for all configurations. In the comparison, the VBM of K and K' points for the P$\uparrow$ and P$\downarrow$ cases are primarily contributed by d$_{z^2}$ and d$_{xz}$/d$_{yz}$ orbitals, respectively. To further understand the electronic properties of Sc$_2$CO$_2$ different polarization states, we investigated the 3D charge density difference (see Fig. 4(b, f)) and plane averaged charge density difference (see Fig. 4(d, h)). Clearly, the charge transfer occurs from Mn$_2$P$_2$S$_3$Se$_3$ to Sc$_2$CO$_2$ for the Sc$_2$CO$_2$ P$\uparrow$ case. On the contrary, when the polarization of Sc$_2$CO$_2$ transforms from P$\uparrow$ to P$\downarrow$, the charge transfer also was switched from Sc$_2$CO$_2$ to Mn$_2$P$_2$S$_3$Se$_3$. Furthermore, we also investigated the band alignments of Mn$_2$P$_2$S$_3$Se$_3$/Sc$_2$CO$_2$ heterostructure, as shown in Fig. 4(c, g). The VBM of Mn$_2$P$_2$S$_3$Se$_3$ is higher than the CBM of Sc$_2$CO$_2$, leading to the charge transfer from Mn$_2$P$_2$S$_3$Se$_3$ to Sc$_2$CO$_2$ for the Sc$_2$CO$_2$ P$\uparrow$ case. When the polarization is switched from P$\uparrow$ to P$\downarrow$, the VBM of Sc$_2$CO$_2$ will be slightly above the VBM of Mn$_2$P$_2$S$_3$Se$_3$, resulting in the direction of charge transfer is also transformed. These results also mutually confirm the correctness of our calculation.

\begin{figure}[htb]
\begin{center}
\includegraphics[angle=0,width=1.0\linewidth]{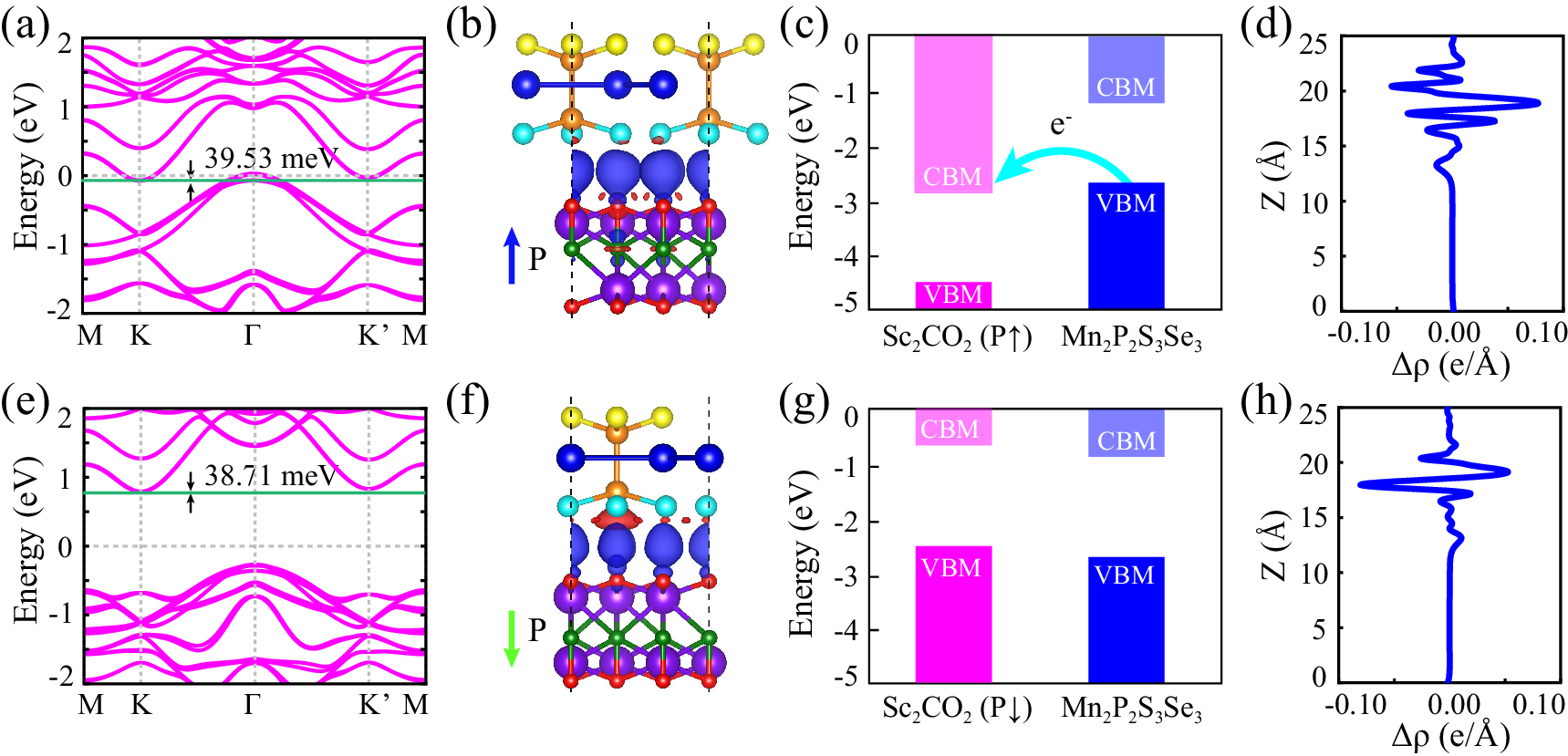}
\caption{The top-O configuration Mn$_2$P$_2$S$_3$Se$_3$/Sc$_2$CO$_2$ heterostructure for the Se-surface. (a, e) Band structures with considering the SOC under the P$\uparrow$ (a) and the P$\downarrow$ (e) states. (b, f) The 3D charge density difference under the P$\uparrow$ (b) and the P$\downarrow$ (f) states. (c, g) Band alignments of the P$\uparrow$ (c) and the P$\downarrow$ (g) states. (d, h) Plane averaged charge density difference for the P$\uparrow$ (d) and the P$\downarrow$ (h) states. The conduction band valley splitting is indicated by the green shading.
}
\end{center}
\end{figure}

\subsection{The nonlinear Hall effect and electronic property detection devices}
Since the nonlinear Hall voltage is proportional to the BCD, however, the BCD is highly correlated with the symmetry. Monolayer Mn$_2$P$_2$S$_3$Se$_3$ has neither the $\emph{P}$ nor $\emph{T}$ symmetry, we analyze the BCD based on these two symmetries. First of all, we analyze the symmetry of Berry curvature $\Omega_{n,d}$, where n is the band index. For the $\emph{P}$ symmetry, $\Omega_{n,d}$ is even, i.e., $\Omega_{n,d}$($\textbf{k}$) = $\Omega_{n,d}$(-$\textbf{k}$). Contrarily, $\Omega_{n,d}$ is odd under the $\emph{T}$ symmetry, i.e., $\Omega_{n,d}$($\textbf{k}$) = -$\Omega_{n,d}$(-$\textbf{k}$). Hence, there are non-zero BCD components in monolayer Mn$_2$P$_2$S$_3$Se$_3$.

The BCD is a 3$\times$3 tensor, D$_{bd}$ is defined as
\begin{equation}
\begin{split}
D_{bd} &= \sum_{n} \int_k  f_n (\textbf{k})\frac{\partial \Omega_{n,d}(\textbf{k})}{\partial \textbf{k}_b}
     \\&= \sum_{n} \int_k \frac{\partial \epsilon_{kn}}{\partial \textbf{k}_b} \Omega_{n,d}(\textbf{k}) \frac{\partial f_n(\textbf{k})}{\partial \epsilon_{kn}}
\end{split}
\end{equation}
where $\emph{f}$$_n$$(\textbf{k})$ is the Fermi-Dirac distribution, and $\Omega_{n,d}(\textbf{k})$ is the Berry curvature. In 2D materials, only the z component of Berry curvature is retained, namely, $\Omega_{n,z}(\textbf{k})$,
\begin{equation}
\Omega_{n,z}(\textbf{k})=-\sum_{n\prime \neq n}\frac{2\rm Im \left \langle \psi_{nk} \mid v_{x} \mid \psi_{n\prime k} \right \rangle \left \langle \psi_{n\prime k} \mid v_{y} \mid \psi_{nk} \right \rangle}{(E_{n\prime}-E_{n})^2},
\end{equation}

The BCD is a pesudo-tensor and determined by \cite{25}
\begin{equation}
\textbf{D} = det(S)S\textbf{D}S^{-1}
\end{equation}
where S represents the symmetric operation matrix of point group. The magnetic point group of monolayer Mn$_2$P$_2$S$_3$Se$_3$ is 3m. The magnetic point group 3m includes a threefold rotation C$_3$ symmetry along z axis (C$_{3z}$), and three mirror reflection perpendicular to xy plane (M$_{xy}$), x axis (M$_{x}$), and, y axis (M$_y$), respectively. The symmetric operator of C$_{3z}$, M$_{xy}$, M$_{x}$, and M$_y$ are
\begin{equation}
  C_{3z}  = \left(
  \begin{array}{cccccccc}
  cos(2\pi/3)      & -sin(2\pi/3)     & 0       \\
  sin(2\pi/3)      &  cos(2\pi/3)     & 0       \\
  0                &  0               & 1       \\
  \end{array}
  \right)
\end{equation}
\begin{equation}
  M_{xy}  = \left(
  \begin{array}{cccccccc}
  0      & -1     & 0       \\
 -1      &  0     & 0       \\
  0      &  0     & 1       \\
  \end{array}
  \right)
\end{equation}
\begin{equation}
  M_{x}  = \left(
  \begin{array}{cccccccc}
 -1      &  1     & 0       \\
  0      &  1     & 0       \\
  0      &  0     & 1       \\
  \end{array}
  \right)
\end{equation}
\begin{equation}
  M_{y}  = \left(
  \begin{array}{cccccccc}
  1      &  0     & 0       \\
  1      & -1     & 0       \\
  0      &  0     & 1       \\
  \end{array}
  \right)
\end{equation}
According to Eqs (7)-(11), it is worth noting that only the D$_{xz}$ and D$_{yz}$ component can exist for monolayer Mn$_2$P$_2$S$_3$Se$_3$ with the 3m magnetic point group \cite{30,46}. As shown in Fig. S9 \cite{43}, it exhibits the calculated BCD as a function of energy for monolayer Mn$_2$P$_2$S$_3$Se$_3$. Obviously, the D$_{xz}$ and D$_{yz}$ is very small since the band crossing is almost non-existent in the -2 to 2 eV energy range.

\begin{figure}[htb]
\begin{center}
\includegraphics[angle=0,width=1.0\linewidth]{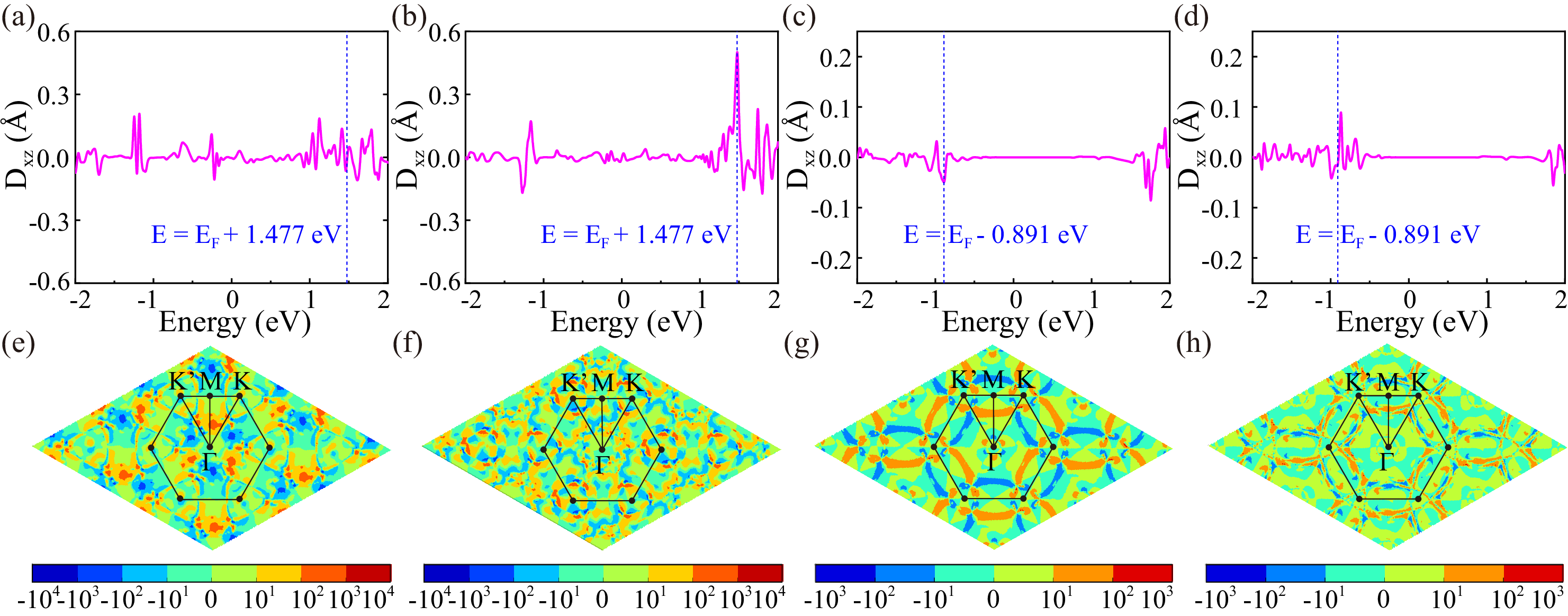}
\caption{(a-d) The calculated BCD components D$_{xz}$ of top-O configurations as a function of energy for the (a)  S-surface Mn$_2$P$_2$S$_3$Se$_3$/Sc$_2$CO$_2$ heterostructure P$\uparrow$, (b) Se-surface Mn$_2$P$_2$S$_3$Se$_3$/Sc$_2$CO$_2$ heterostructure P$\uparrow$, (c) S-surface Mn$_2$P$_2$S$_3$Se$_3$/Sc$_2$CO$_2$ heterostructure P$\downarrow$, and (d) Se-surface Mn$_2$P$_2$S$_3$Se$_3$/Sc$_2$CO$_2$ heterostructure P$\downarrow$. (e-h) The 2D $\emph{k}$-resolved Berry curvature $\Omega_z$ for the (e) S-surface Mn$_2$P$_2$S$_3$Se$_3$/Sc$_2$CO$_2$ heterostructure P$\uparrow$ at E$_{\rm F}$ + 1.477 eV, (f) Se-surface Mn$_2$P$_2$S$_3$Se$_3$/Sc$_2$CO$_2$ heterostructure P$\uparrow$ at E$_{\rm F}$ + 1.477 eV, (g) S-surface Mn$_2$P$_2$S$_3$Se$_3$/Sc$_2$CO$_2$ heterostructure P$\downarrow$ at E$_{\rm F}$ - 0.891 eV, and (h) Se-surface Mn$_2$P$_2$S$_3$Se$_3$/Sc$_2$CO$_2$ heterostructure P$\downarrow$ at E$_{\rm F}$ - 0.891 eV.
}
\end{center}
\end{figure}

\begin{figure}[htb]
\begin{center}
\includegraphics[angle=0,width=1.0\linewidth]{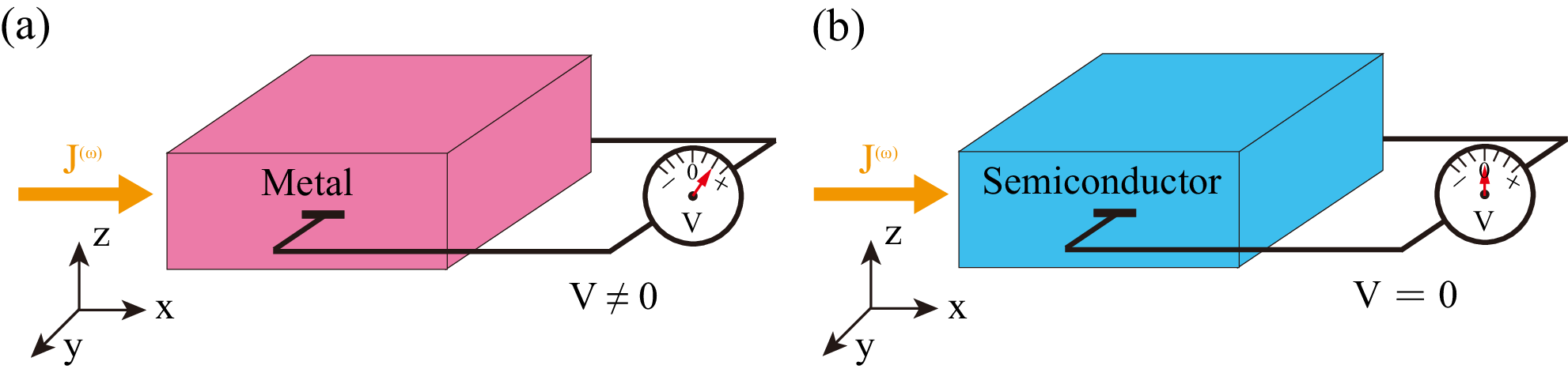}
\caption{Schematics of the electronic property detection device by the NHE. The driving current flows along the x axis, the nonlinear Hall voltage generates along the y axis, which is determined by the electronic property. (a) The nonlinear Hall voltage is (a) non-zero for the valley polarized metal, and (b) zero for the valley polarized semiconductor.
}
\end{center}
\end{figure}

When the Mn$_2$P$_2$S$_3$Se$_3$ and Sc$_2$CO$_2$ form the heterostructure, the point group decreases from C$_{3v}$ to C$_{1h}$. Consequently, only the M$_y$ mirror symmetry is preserved. Therefore, the BCD component D$_{yz}$ disappears, while the D$_{xz}$ is still retained. Figure 5 shows the calculated BCD components D$_{xz}$ for top-O configuration Mn$_2$P$_2$S$_3$Se$_3$/Sc$_2$CO$_2$ heterostructure. For the P$\uparrow$ case (see Fig. 5(a, b)), D$_{xz}$ has a non-zero value at the Fermi level. When the P switches $\downarrow$, the D$_{xz}$ becomes zero. It arises from whether there is a band gap in the two polarization states. In addition, the D$_{xz}$ of P$\uparrow$ is remarkably larger than that of P$\downarrow$. Obviously, the value of D$_{xz}$ reached 0.50 $\rm \AA$ at E$_{\rm F}$ + 1.477 eV for Se-surface. To understand its origin, we calculated the corresponding k-resolved Berry curvatures for the P$\uparrow$, as shown in Fig. 5(e, f). It can clearly be seen that the contribution near the $\Gamma$ point is very large. For the S-surface, the positive and negative Berry curvatures are almost equivalent, so the D$_{xz}$ is smaller. While the Se-surface is predominantly positive Berry curvature near the $\Gamma$ point. Similarly, the k-resolved Berry curvatures are investigated for the P$\downarrow$ at E$_{\rm F}$ - 0.891 eV. Different from the P$\uparrow$ case, the main contribution comes from the vicinity of K and K' points. And the S-surface is significantly larger than the Se-surface, which is consistent with D$_{xz}$.

The manipulation and detection of charge, spin, or valley is the basis of realizing electronic, spintronic, and valleytronic devices. Similarly, The valley polarized metal-semiconductor transition also can be detected and manipulated using the valleytronic devices. Our calculation results show that the NHE can effectively reflect the valley polarized metal-semiconductor phase transition. If the D$_{xz}$ is non-zero for the valley polarized metal, in the Mn$_2$P$_2$S$_3$Se$_3$/Sc$_2$CO$_2$ heterostructure P$\uparrow$, a non-zero Hall signal will be detected along the y axis, as shown in Fig. 6(a). On the contrary, when the ferroelectric polarization of Sc$_2$CO$_2$ switches to P$\downarrow$, the valley polarized semiconductor cannot generate the nonlinear Hall voltage due to no bands near the Fermi level (see Fig. 6(b)). The detection and manipulation of valley polarized metal-semiconductor transition will be conducive to the flourishing of electronics, spintronics, and valleytronics.

\section{CONCLUSION}
In summary, we systematically investigated the stability, magnetic ground state, electronic properties, and NHE for vdW Mn$_2$P$_2$S$_3$Se$_3$/Sc$_2$CO$_2$ heterostructures. We find that monolayer Mn$_2$P$_2$S$_3$Se$_3$ is a robust AFM valley semiconductor for the Coulomb repulsion U. After forming heterostructure with Sc$_2$CO$_2$, the valley splitting is further enhanced. Importantly, when the Mn$_2$P$_2$S$_3$Se$_3$/Sc$_2$CO$_2$ is the P$\uparrow$ state, it exhibits a valley polarization metallic property. However, while the ferroelectric polarization of Sc$_2$CO$_2$ transforms from P$\uparrow$ to P$\downarrow$, the system will also be accompanied by the realization phase transition from valley polarized metal to valley polarized semiconductor. It is worth noting that this phenomenon exists on both the S-surface and Se-surface. Moreover, we also find that the BCD can be effectively tuned by switching Sc$_2$CO$_2$ polarization direction. Based on this, we propose that the NHE devices can be employed to detect valley polarized metal-semiconductor transitions. Our findings are expected to broaden the application scenarios of the NHE.

\section*{ACKNOWLEDGEMENTS}
This work is supported by the National Natural Science Foundation of China (Grants No. 12474238, and No. 12004295). P. Li also acknowledge supports from the China's Postdoctoral Science Foundation funded project (Grant No. 2022M722547), the Fundamental Research Funds for the Central Universities (xxj03202205), the Open Project of State Key Laboratory of Surface Physics (No. KF2024$\_$02), and the Open Project of State Key Laboratory of Silicon and Advanced Semiconductor Materials (No. SKL2024-10).



\begin{thebibliography}{99}

\bibitem{1} G. E. Moore, Cramming more components onto integrated circuits, Proc. IEEE 86, 82 (1998).

\bibitem{2} A. D. Franklin, Nanomaterials in transistors: From high-performance to thin-film applications, Science 349, aab2759 (2015).

\bibitem{3} M. Wang, S. Cai, C. Pan, C. Wang, X. Lian, Y. Zhuo, K. Xu, T. Cao, X. Pan, B. Wang, S. J. Liang, J. J. Yang, P. Wang, and F. Miao, Robust memristors based on layered two-dimensional materials, Nat. Electron. 1, 130 (2018).

\bibitem{4} X. Liu, and M. C. Hersam, 2D materials for quantum information science, Nat. Rev. Mater. 4, 669 (2019).

\bibitem{5} P. Li, X. S. Zhou, and Z. X. Guo, Intriguing Magnetoelectric Effect in Two-dimensional Ferromagnetic/Perovskite Oxide Ferroelectric Heterostructure. npj Comput. Mater. 8, 20 (2022).

\bibitem{6} P. Li, J. Z. Zhang, Z. X. Guo, T. Min, and X. R. Wang, Intrinsic anomalous spin Hall effect. Sci. China Phys. Mech. 66, 227511 (2023).

\bibitem{7} A. Zhang, Z. Gong, Z. Zhu, A. Pan, and M. Chen, Effects of the substrate-surface reconstruction and orientation on the spin valley polarization in MoTe$_2$/EuO, Phys. Rev. B 102, 155413 (2020).

\bibitem{8} Y. Liu, X. Duan, H. Shin, S. Park, Y. Huang, and X. Duan, Promises and prospects of two-dimensional transistors, Nature 591, 43 (2021).

\bibitem{9} R. Quhe, Z. Di, J. Zhang, Y. Sun, L. Zhang, Y. Guo, S. Wang, and P. Zhou, Asymmetric conducting route and potential redistribution determine the polarization-dependent conductivity in layered ferroelectrics, Nat. Nanotechnol. 19, 173 (2024).

\bibitem{10} T. Latychevskaia, D. A. Bandurin, and K. S. Novoselov, A new family of septuple-layer 2D materials of MoSi$_2$N$_4$-like crystals, Nat. Rev. Phys. 6, 426 (2024).

\bibitem{11} W. Y. Tong, S. J. Gong, X. Wan, and C. G. Duan, Concepts of ferrovalley material and anomalous valley Hall effect, Nat. Commun. 7, 13612 (2016).

\bibitem{12} P. Li, C. Wu, C. Peng, M. Yang, and W. Xun, Multifield tunable valley splitting in two-dimensional MXene Cr$_2$COOH, Phys. Rev. B 108, 195424 (2023).

\bibitem{13} Y. Liu, Y. Feng, Y. Dai, B. Huang, and Y. Ma, Engineering Layertronics in Two-Dimensional Ferromagnetic Multiferroic Lattice, Nano Lett. 24, 3507 (2024).

\bibitem{14} P. Li, B. Liu, S. Chen, W. X. Zhang, and Z. X. Guo, Progress on two-dimensional ferovalley materials, Chin. Phys. B 33, 017505 (2024).

\bibitem{15} J. Zhao, Y. Feng, Y. Dai, B. Huang, and Y. Ma, Ferroelectrovalley in Two-Dimensional Multiferroic Lattices, Nano Lett. 24, 10490 (2024).

\bibitem{16} G. Yu, J. Ji, C. Xu, and H. J. Xiang, Bilayer stacking ferrovalley materials without breaking time-reversal and spatial-inversion symmetry, Phys. Rev. B 109, 075434 (2024).

\bibitem{17} D. MacNeill, C. Heikes, K. F. Mak, Z. Anderson, A. Kormanyos, V. Zolyomi, J. Park, and D. C. Ralph, Breaking of Valley Degeneracy by Magnetic Field in Monolayer MoSe$_2$, Phys. Rev. Lett. 114, 037401 (2015).

\bibitem{18} M. U. Rehman, Z. U. Rahman, and M. Kiani, Emerging spintronic and valleytronic phenomena in noncentrosymmetric variants of the Kane-Mele X$_4$Y$_2$Z$_6$ materials family (X = Pt, Pd, Ni; Y = Hg, Zn, Cd; Z = S, Se, Te), Phys. Rev. B 109, 165424 (2024).

\bibitem{19} W. Xun, C. Wu, H. Sun, W. Zhang, Y. Z. Wu, and P. Li, Coexisting Magnetism, Ferroelectric, and Ferrovalley Multiferroic in Stacking-Dependent Two-Dimensional Materials, Nano Lett. 24, 3541 (2024).

\bibitem{20} K. Shao, H. Geng, E. Liu, J. L. Lado, W. Chen, and D. Y. Xing, Non-Hermitian Moire Valley Filter, Phys. Rev. Lett. 132, 156301 (2024).

\bibitem{21} H. Hu, W. Y. Tong, Y. H. Shen, X. Wan, C. G. Duan, Concepts of the half-valley-metal and quantum anomalous valley Hall effect, npj Conput. Mater. 6, 129 (2020).

\bibitem{22} W. Pan, Tuning the magnetic anisotropy and topological phase with electronic correlation in single-layer H-FeBr$_2$, Phys. Rev. B 106, 125122 (2022).

\bibitem{23} Z. He, R. Peng, X. Feng, X, Xu, Y. Dai, B. Huang, and Y. Ma, Two-dimensional valleytronic semiconductor with spontaneous spin and valley polarization in single-layer Cr$_2$Se$_3$, Phys. Rev. B 104, 075105 (2021).

\bibitem{24} K. Wang, Y. Li, H. Mei, P. Li, and Z. X. Guo, Quantum anomalous Hall and valley quantum anomalous Hall effects in two-dimensional d$^0$ orbital XY monolayers, Phys. Rev. Mater. 6, 044202 (2022).

\bibitem{25} Y. Wu, J. Tong, L. Deng, F. Luo, F. Tian, G. Qin, and X. Zhang, Coexisting Ferroelectric and Ferrovalley Polarizations in Bilayer Stacked Magnetic Semiconductors, Nano Lett. 23, 6226 (2023).

\bibitem{26} P. Li, X. Yang, Q. S. Jiang, Y. Z. Wu, and W. Xun, Built-in electric field and strain tunable valley-related multiple topological phase transitions in VSiXN$_4$(X=C,Si,Ge,Sn,Pb) monolayers, Phys. Rev. Mater. 7, 064002 (2023).

\bibitem{27} S. D. Guo, L. Zhang, Y. Zhang, P. Li, and G. Wang, Large spontaneous valley polarization and anomalous valley Hall effect in antiferromagnetic monolayer Fe$_2$CF$_2$, Phys. Rev. B 110, 024416 (2024).

\bibitem{28} I. Sodemann, and L. Fu, Quantum Nonlinear Hall Effect Induced by Berry Curvature Dipole in Time-Reversal Invariant Materials, Phys. Rev. Lett. 115, 216806 (2015).

\bibitem{29} T. Low, Y. Jiang, and F. Guinea, Topological currents in black phosphorus with broken inversion symmetry, Phys. Rev. B 92, 235447 (2015).

\bibitem{30} R. C. Xiao, D. F. Shao, Z. Q. Zhang, and H. Jiang, Two-Dimensional Metals for Piezoelectriclike Devices Based on Berry-Curvature Dipole, Phys. Rev. Appl. 13, 044014 (2020).

\bibitem{31} N. N. Zhao, Z. F. Ouyang, P. H. Sun, J. F. Zhang, K. Liu, and Z. Y. Lu, Nonlinear Hall effect and potential Ising superconductivity in monolayer MXene heterostructure of T-Mo$_2$C/H-Mo$_2$C, Phys. Rev. B 108, 035140 (2023).

\bibitem{32} C. Wang, Y. Gao, and D. Xiao, Intrinsic Nonlinear Hall Effect in Antiferromagnetic Tetragonal CuMnAs, Phys. Rev. Lett. 127, 277201 (2021).

\bibitem{33} D. Kaplan, T. Holder, and B. Yan, General nonlinear Hall current in magnetic insulators beyond the quantum anomalous Hall effect, Nat. Commun. 14, 3053 (2023).

\bibitem{34} P. E. Blochl, Projector augmented-wave method, Phys. Rev. B 50, 17953 (1994).

\bibitem{35} G. Kresse, and J. Furthmuller, Efficient iterative schemes for ab initio total-energy calculations using a plane-wave basis set, Phys. Rev. B 54, 11169 (1996).

\bibitem{36} J. P. Perdew, K. Burke, and M. Ernzerhof, Generalized Gradient Approximation Made Simple, Phys. Rev. Lett. 77, 3865 (1996).

\bibitem{37} S. Grimme, J. Antony, S. Ehrlich, and H. Krieg, A consistent and accurate ab initio parameterization of density functional dispersion correction (DFT-D) for the 94 elemenets H-Pu, J. Chem. Phys. 132, 154104 (2010).

\bibitem{38} A. Togo, and I. Tanaka, First principles phonon calculations in materials science, Scr. Mater. 108, 1 (2015).

\bibitem{39} A. A. Mostofi, J. R. Yates, Y. S. Lee, I. Souza, D. Vanderbilt, and N. Marzari, wannier90: A tool for obtaining maximally-localised Wannier functions, Commput. Phys. Commun. 178, 685 (2008).

\bibitem{40} A. A. Mostofi, J. R. Yates, G. Pizzi, Y. S. Lee, I. Souza, D. Vanderbilt, and N. Marizari, An updated version of wannier90: A tool for obtaining maximally-localised Wannier functions, Commput. Phys. Commun. 185, 2309 (2014).

\bibitem{41} P. Li, X. Li, J. Feng, J. Ni, Z. X. Guo, and H. Xiang, Origin of zigzag antiferromagnetic orders in XPS$_3$ (X = Fe, Ni) monolayers, Phys. Rev. B 109, 214418 (2024).

\bibitem{42} A. Chandrasekaran, A. Mishra, and A. K. Singh, Ferroelectricity, antiferroelectricity, and ultrathin 2D electron/hole gas in multifunctional monolayer MXene, Nano Lett. 17, 3290 (2017).

\bibitem{43} See Supplemental Material at http://link.aps.org/supplemental/xxx for the band structure of monoalyer Mn$_2$P$_2$S$_3$Se$_3$ with considering the SOC, the Mn$_2$P$_2$S$_3$Se$_3$/Sc$_2$CO$_2$ heterostructures configurations for Se-surface, spin-polarized band structures of Mn$_2$P$_2$S$_3$Se$_3$/Sc$_2$CO$_2$ heterostructures for the S-surface, the band structure of Mn$_2$P$_2$S$_3$Se$_3$/Sc$_2$CO$_2$ heterostructures for the S-surface with considering the SOC, the projected band structures of Mn$_2$P$_2$S$_3$Se$_3$/Sc$_2$CO$_2$ heterostructures for the S-surface with considering the SOC, spin-polarized band structures of Mn$_2$P$_2$S$_3$Se$_3$/Sc$_2$CO$_2$ heterostructures for the Se-surface, the band structures of Mn$_2$P$_2$S$_3$Se$_3$/Sc$_2$CO$_2$ heterostructures for the Se-surface with considering the SOC, the projected band structures of Mn$_2$P$_2$S$_3$Se$_3$/Sc$_2$CO$_2$ heterostructures for the Se-surface with considering the SOC, the calculated BCD components $\emph{D}_{xz}$ and $\emph{D}_{yz}$ as a function of energy for monolayer Mn$_2$P$_2$S$_3$Se$_3$, the energy difference between the ferromagnetic and antiferromagnetic states and the MAE for monolayer Mn$_2$P$_2$S$_3$Se$_3$, the energy difference between the ferromagnetic and antiferromagnetic states for different stacking configurations Mn$_2$P$_2$S$_3$Se$_3$/Sc$_2$CO$_2$ heterostructures, and the MAE of different stacking configurations for Mn$_2$P$_2$S$_3$Se$_3$/Sc$_2$CO$_2$ heterostructures.

\bibitem{44} P. Li, and T. Y. Cai, Two-Dimensional Transition-Metal Oxides Mn$_2$O$_3$ Realized the Quantum Anomalous Hall effect, J. Phys. Chem. C 124, 12705 (2020).

\bibitem{45} W. Du, R. Peng, Z. He, Y. Dai, B. Huang, and Y. Ma, Anomalous valley Hall effect in antiferromagnetic monolayers, npj 2D Mater. Appl. 6, 11 (2020).

\bibitem{46} C. Wang, Y. Gao, and D. Xiao, Intrinsic Nonlinear Hall Effect in Antiferromagnetic Tetragonal CuMnAs, Phys. Rev. Lett. 127, 277201 (2021).








	
\end{thebibliography}
\end{document}